\documentstyle[11pt,aaspp4]{article}

\slugcomment{ApJ Letters (in press)}

\lefthead{M. S. Sahu et al.}
\righthead{The D/H Ratio in Interstellar Gas}

\begin{document}

\title{The D/H Ratio in Interstellar Gas Towards G191-B2B\altaffilmark{1}}

\author{M. S. Sahu\altaffilmark{2,3}, W. Landsman\altaffilmark{4},
F. C. Bruhweiler\altaffilmark{5}, T. R. Gull\altaffilmark{2}, C. A. Bowers\altaffilmark{2}, D. Lindler\altaffilmark{6}, K. Feggans\altaffilmark{6},
M. A. Barstow\altaffilmark{7}, I. Hubeny\altaffilmark{2,3}, J. B. Holberg\altaffilmark{8}}

\altaffiltext{2}{NASA/Goddard Space Flight Center, Code 681, Greenbelt, MD~20771}
\altaffiltext{3}{National Optical Astronomy Observatories, 950 North Cherry Avenue,Tucson, AZ~87519-4933}
\altaffiltext{4}{Raytheon ITSS, NASA/Goddard Space Flight Center, Code 681, Greenbelt, MD~20771}
\altaffiltext{5}{IACS/Department of Physics, Catholic University of America, Washington DC~20064}
\altaffiltext{6}{Advanced Computer Concepts, Inc./Goddard Space Flight Center
Code 681, Greenbelt, MD~20771}
\altaffiltext{7}{Dept. of Physics \& Astronomy, University of Leicester, Leicester,
LE1 RH, UK}
\altaffiltext{8}{University of Arizona, Lunar and Planetary Lab., West 
Gould-Simpson Bldg. Rm. 901, Tucson, AZ~85721}

\altaffiltext{1}{Based on observations done with the NASA/ESA {\it
Hubble Space Telescope} obtained at the Space Telescope Science Institute,
which is operated by the Association of Universities for Research in Astronomy
(AURA), Inc. under NASA contract NAS5-26555}

\begin{abstract}

Recent analysis of Goddard High Resolution Spectrograph (GHRS) echelle spectra suggests $\sim$ 30\%
variations in the D/H abundance ratio along the line-of-sight to the nearby (69~pc) hot, white dwarf (WD) G191-B2B (Vidal-Madjar et al. 1998, hereafter VM98).   Variations in the D/H ratio on such short length scales imply non-uniform production/destruction of deuterium and an inefficient mixing of gas in the local interstellar medium (LISM).
We reinvestigate the question of spatial variation of the  local D/H
abundance, using both archival GHRS spectra, and new echelle spectra of G191-B2B
obtained with the Space Telescope Imaging Spectrograph (STIS) aboard 
HST. The STIS spectra were obtained in the high-resolution (E140H) mode
and  cover the wavelength region 1140 to 1700 \AA. Our
analysis uses stratified line-blanketed non-LTE model atmosphere
calculations to determine the shape of the intrinsic WD Lyman-$\alpha$ profile
and estimate the WD photospheric contamination of the interstellar lines.    Although three velocity components were reported previously towards
G191-B2B, we detect only two velocity components. The first component is at $v$$_{hel}$ $\sim$ 8.6 km s$^{-1}$ and the second at $v$$_{hel}$ $\sim$ 19.3 km s$^{-1}$, which we identify with the Local Interstellar Cloud (LIC).   From the STIS data we derive  D/H =
1.60$\pm$$\stackrel{0.39}{_{0.27}}$$\times$10$^{-5}$ for the LIC component, and
D/H $>$ 1.26$\times$10$^{-5}$ for the 8.6 km s$^{-1}$ component (uncertainties denote 2$\sigma$ or 95\% confidence limits). The derived D/H  values in both
components are  consistent with (D/H)$_{LIC}$ =
1.5$\pm$0.1$\times$10$^{-5}$ determined by 
Linsky (1998). The STIS data  provide no evidence for local or
component-to-component variation in the D/H ratio.   

Our re-analysis of the GHRS data gives essentially the same
results as VM98, despite using two velocity components for the profile fitting
(versus three by VM98) and using a more physically realistic WD
Lyman-$\alpha$ profile for G191-B2B.   The GHRS data indicate a
component-to-component variation as well as a variation of the D/H ratio in the
LISM, neither of which are supported by the newer STIS data. 
The D~{\sc i} absorption in the GHRS spectrum is shallower than in the STIS spectrum. We believe the most probable cause for this difference in the two data sets is the characterization of the background due to scattered light in the GHRS and STIS spectrographs. The D/H ratios derived are sensitive to the background subtraction procedures employed. The two-dimensional MAMA detectors of STIS measure both the spatial and wavelength dependences of scattered light, allowing more accurate scattered light corrections than was possible with GHRS.

\end{abstract}


\keywords{white dwarfs (G191-B2B) --- ISM: abundances --- ultraviolet:ISM}


%

\section{Introduction}

The best post-IUE measurement of the D/H ratio in the LISM within $\sim$100~pc has been the HST-GHRS study by Linsky et al. 
(L93 \& L95), who found D/H = 1.6 $\pm$ 0.2$\times$ 10$^{-5}$ towards 
Capella, a nearby late-type spectroscopic binary system [$d$ = 12.5~pc; ($l$,$b$) = (162$^{\circ}$.6, $+$4$^{\circ}$.6)].
Subsequent GHRS measurements towards other nearby late-type stars and WDs, 
reported by various authors indicate that within 
measurement uncertainties, the D/H ratio in the LISM is constant and the value is consistent with Linsky et al.'s results (Landsman et al. 1996, Piskunov et al. 1997, Dring et al. 1998). 
Only one result suggests significant variation of the D/H ratio within 69~pc. Using GHRS echelle data, VM98 reported the presence of three 
velocity components towards the white dwarf G191-B2B, with D/H ratios ranging between 
0.9$\times$10$^{-5}$ and 1.56$\times$10$^{-5}$ implying a variation of the D/H ratio 
by $\sim$ 30\% in the LISM. Towards the more distant 
star $\delta
$Ori~A ($d$ $\sim$ 350~pc), however, Interstellar Medium Absorption Profile
Spectrograph (IMAPS) observations by Jenkins et al. (1999), suggest 
spatial variations of the D/H ratio.  Current models of Galactic chemical evolution predict variations in the D/H ratio over length scales of $\sim$ 1~kpc (e.g. Tosi, 1998) 
but not over length scales as short as 69~pc. Possible implications of the spatial variation of the D/H ratio
over such short length scales include a non-primordial source of
deuterium production (Mullan \& Linsky 1998). We have reinvestigated the question of spatial variation of the D/H ratio towards G191-B2B, using newer HST-Space Telescope Imaging Spectrograph (STIS) data. Compared to the GHRS echelle data, the STIS data have better scattered light corrections and include lines such as 
Si~{\sc ii} ($\lambda$$\lambda$1190, 1193, 1260 and 1526) and Fe~{\sc ii} $\lambda$1608.5  which were not observed with GHRS.
In this Letter, we report the D/H ratio derived from STIS data and the disagreement between the D/H ratios derived from STIS and GHRS data . The most probable reason for this disagreement is the inadequate scattered light corrections available for the GHRS data ($\S$6). An extended analysis of the 
STIS and GHRS data will be presented elsewhere.
\section{Observations and data reduction}
G191-B2B was observed with STIS on 1998 December 17 
using the high-resolution E140H (R$\sim$110,000) mode in the $\sim$ 1140 to 1700\AA\ region. These observations were part of STIS flux calibrations and the entrance aperture used was  0.2 $\times$ 0.2 arcsec. The use of this aperture decreases the effective spectral resolution, particularly in the bluer wavelength regions near the Lyman-$\alpha$ region (where the telescope PSF halo is more pronounced). The spectra were processed with the IDL-based CALSTIS reduction package developed
by the STIS Instrument Development Team (IDT) at Goddard Space Flight Center
(Lindler 1999). There was appreciable scattered light in the Lyman-$\alpha$ region ($\sim$ 11\% of the continuum at 1213\AA) which was corrected using an iterative correction algorithm developed by the STIS IDT (Bowers \& Lindler 1999, in preparation). This algorithm models several sources of stray light including the telescope PSF halo, echelle grating scatter and the detector halo.   Archival GHRS echelle data of the G191-B2B (observed 1995 July 26 to 28) were reduced using the IDL-based CALHRS routine developed by the GHRS IDT at Goddard. Three individual 
GHRS FP-SPLIT exposures are combined to obtain the final spectrum  containing the D~{\sc i} and H~{\sc i} interstellar lines. The core of the saturated Lyman-$\alpha$ absorption dips below zero flux and was corrected by setting the inter-order coefficients $a$ and $b$ at 0.9 (Cardelli et al. 1993).
Near the D~{\sc i} feature, the S/N per data point in the continuum for both GHRS and STIS is $\sim$ 20, although GHRS has better wavelength sampling (0.003\AA\ versus 0.005\AA\ for STIS).
\section{Use of NLTE stellar atmosphere models}
To disentangle the interstellar 
D~{\sc i} and H~{\sc i} absorption from the observed profiles,
it is essential to use the most physically realistic model for the intrinsic stellar Lyman-$\alpha$ profile. G191-B2B  [$d$ = 68.8~pc; ($l$,$b$) = (155$^{\circ}$.9, $+$7$^{\circ}$.1)] (Vauclair et al. 1997), belongs to a class of hot, DA WDs that contain significant amounts
of heavy elements such as C, N, O, Si, Fe and Ni in their atmospheres.
Lanz et al. (1996) performed NLTE  calculations including the effects of line-blanketing from more than 9 $\times$ 10$^6$ atomic transitions (mainly Fe and Ni) and matched the flux level and shape of the EUV spectrum of G191-B2B for the first time. The apparent effective temperature
of WDs like G191-B2B is sensitive to assumptions about the photospheric composition (Barstow et al. 1998) and must be taken into account in 
modeling the stellar Lyman-$\alpha$ profile.
Barstow and co-workers (in preparation) have refined their stratified 
line-blanketed NLTE calculations (Barstow et al. 1999) and the best-fit model atmosphere [T$_{eff}$ = 54,000 $\pm$ 2000~K and log~$g$ = 7.5$\pm$ 0.03] is adopted in our analysis to predict the intrinsic WD Lyman-$\alpha$ profile and to check for contamination of the interstellar lines by narrow WD absorption lines.
The radial velocity of G191-B2B used in our analysis, estimated from STIS data of other WD lines (Bruhweiler et al. 1999, in preparation) is 24.6 $\pm$ 0.4 km~s$^{-1}$ (including gravitational redshift).
\section{The number of velocity components in the line-of-sight}
In addition to the interstellar  D~{\sc i}
and H~{\sc i} absorption lines, the STIS echelle spectra show interstellar absorption due to N~{\sc i}
($\lambda$$\lambda$1199.5, 1200.2 and 1200.7), 
C~{\sc ii} $\lambda$1334.5, 
C$^\ast$~{\sc ii} $\lambda$1335.7, O~{\sc i} $\lambda$1302, 
Si~{\sc ii} ($\lambda$$\lambda$1190, 1193, 1260, 1304 and 1526), Si~{\sc iii} $\lambda$1206.5,
Al~{\sc ii} $\lambda$1670.8, S~{\sc ii} $\lambda$1259.5  and Fe~{\sc ii} $\lambda$1608.5.
The interstellar N~{\sc i} $\lambda$1200.7, Si~{\sc ii} $\lambda$$\lambda$ 1193 \& 1304 and Fe~{\sc ii} lines are not contaminated by WD lines. 
Figure 1 (a, b) shows the profile fits to the  N~{\sc i} $\lambda$ 1200.7
and Si~{\sc ii} $\lambda$ 1304 lines  and two distinct velocity components are seen in these uncontaminated lines (all velocities are in the heliocentric frame).
One component is at $\sim$ 8.6 km~s$^{-1}$ (hereafter referred to as
comp 1). The other component is at 19.3 km~s$^{-1}$, which is within measurement uncertainties of the projected velocity of the LIC (20.3 km~s$^{-1}$) in the line-of-sight to G191-B2B (Lallement et al. 1995). This component is also seen in the Capella data (L93), suggesting both the G191-B2B and Capella sightlines intercept the LIC. Figure 1c shows the STIS and GHRS Si~{\sc iii} 1206.5 \AA\ line utilized by VM98 to determine the number of components in the line-of-sight towards G191-B2B. Our GHRS profile is shifted to lower velocities by $\sim$ 4 km~s$^{-1}$ compared to the profile in the top panel of Figure 5 in VM98. Unlike VM98, we are able to obtain an excellent fit to both the STIS and GHRS profiles of Si~{\sc iii} using only these two components. Detailed profile fitting of the other interstellar species confirmed that no more than two components are required (within the constraints imposed by S/N and spectral resolution of the STIS data) to yield acceptable fits. Our analysis of the STIS and GHRS data explicitly assumes the existence of two distinct components. 
\section{Profile fitting of the interstellar D~{\sc i} and H~{\sc i} lines}
Each component is assumed to be homogeneous and characterized by
a column density $N$, radial velocity $v$ and a line-of-sight 
velocity dispersion defined by $b$ = (2kT/$m$ + $\xi$$^2$)$^{1/2}$
where  $\xi$ is the turbulent velocity parameter along the line-of-sight,
T is the kinetic temperature and $m$ is the ion mass.
The D~{\sc i} and H~{\sc i} interstellar lines were fit simultaneously since they are separated by only 0.33\AA\ and the D~{\sc i} absorption is located on the wing of the broad H~{\sc i} absorption. Line profiles were convolved with either the STIS instrumental LSF for the
0.2 $\times$ 0.2 arcsec slit given by the STIS Handbook (Sahu 1999) or the two-component Gaussian LSF for GHRS given by Spitzer \& Fitzpatrick (1993). The turbulent velocity parameters for the two components were determined by plotting the $b$ values for the various atomic species
as a function of ion mass $m$ and performing a least-squares fit. The best-fit $\xi$ value for the LIC component
is 1.7 km s$^{-1}$ (consistent with L93, L95) while for comp 1, $\xi$ is 2.5 km s$^{-1}$.
The STIS spectra near the Fe~{\sc ii} (the heaviest ion) absorption have low S/N and the $\xi$ value is probably not very accurate. However, the 
derived column densities are insensitive to the assumed values of $\xi$, and the D/H ratios presented here are not affected. 
For modeling of the Lyman-$\alpha$ profile, the velocities of the two components are kept fixed at 8.6 (comp 1) and 19.3 km s$^{-1}$ (LIC) and the $\xi$ values are fixed at 1.7 (LIC) and 2.5 km s$^{-1}$ (comp 1). Two types of profile fits are done for the STIS and GHRS data sets:
(1) keeping the value of D/H free in both components (STIS-FREE,
GHRS-FREE) and (2) forcing the same value of D/H in both components
(STIS-FIXED and GHRS-FIXED). Table 1 lists the results of the profile fitting for the STIS and GHRS data sets. The parameters obtained for the two components are the H~{\sc i} column density, N(H~{\sc i}), the temperature derived from the H~{\sc i} thermal velocity dispersion and the D/H ratio (listed for 
comp 1 and the LIC component respectively in columns 4 through 9). The uncertainties quoted for the D/H ratios denote the 2$\sigma$ (95\%) confidence limits obtained using the method of constant $\chi$$^2$ boundaries (Press et al. 1992). Figure 2 (a, b) shows the best-fit models to the STIS and GHRS data 
for the STIS-FREE and GHRS-FREE cases respectively. The total H~{\sc i} column density towards G191-B2B obtained from the STIS data 
is $\sim$2.04$\times$10$^{18}$ cm$^{-2}$, consistent with the value of 2.05$\times$10$^{18}$ cm$^{-2}$ derived from
the best-fit parameters to the EUVE data over the wavelength range
100 to 500\AA\ (Barstow et al. 1999). 
The D/H ratios derived for the two components from the STIS and GHRS data clearly disagree (compare STIS-FREE and GHRS-FREE cases in Table 1). 
\section {Why do STIS and GHRS data give different values of the D/H ratios?}
Figure 3 compares the STIS and GHRS spectra in the region of the D~{\sc
i} absorption. Note that the D~{\sc i} absorption in the GHRS spectrum is
shallower than in the STIS spectrum. The difference in the derived D/H  values is unlikely to be due to statistical fluctuations. For example, a value of D/H = 1.35 $\times 10^{-5}$ in both components is 3$\sigma$ {\em above} the best-fit determination from the GHRS-FIXED  fit, and 3$\sigma$ {\em below} the best-fit determination from the STIS-FIXED fit.  Another unlikely possibility for the difference in two data sets is time variability in the observed profile near the D~{\sc i} feature, perhaps due to a stellar wind. While Barstow 
et al. (1999) do suggest the presence of a weak stellar wind in G191-B2B to maintain the stratification of the Fe abundances,  they point out that the wind must be less than $10^{-16}$ M$_{\odot}$/yr to avoid elimination of the heavy elements in the photosphere. Such a weak wind would not be detectable, even in Lyman-$\alpha$. The most probable cause  is a systematic error in one of the data sets and two lines of evidence suggest that this error is
more likely to be in the GHRS data. First, whereas $\chi^2$/$\nu$ value of
1.065 for the model fit to the STIS data indicates a 16\% probability that the model is correct and that the uncertainties are correctly estimated, the $\chi^2$/$\nu$ value of 1.200 for the GHRS data indicates only a 0.01\% probability of this being true. Second, when the D/H ratio is kept fixed in both components (STIS-FIXED and GHRS-FIXED), the D/H ratios derived with the STIS data (1.71$\pm$$\stackrel{0.32}{_{0.24}}$$\times$10$^{-5}$) are consistent with the value of 1.5$\pm$0.1$\times$10$^{-5}$ determined for the LIC by Linsky
(1998).  The corresponding value derived for the GHRS data
(1.17$\pm$$\stackrel{0.12}{_{0.11}}$$\times$10$^{-5}$) is not consistent with
observed LIC values. 
Fixed-pattern noise or wavelength drifts during FP-SPLIT subexposures in the GHRS data set could result in a shallower D~{\sc i} absorption. However, the observed scatter in the flux level among the 144 subexposures (comprising the three FP-SPLIT GHRS exposures), shows good agreement with the errors predicted by the CALHRS routine, making this  possibility unlikely. We believe the most probable cause for the difference in the two data sets around the D~{\sc i} absorption is the characterization of the background due to scattered light in the GHRS and STIS spectrographs.  The
D~{\sc i} line located on the wing of the H~{\sc i} profile, almost reaches
zero flux in the core and is sensitive to the background subtraction
procedures employed.  The two-dimensional MAMA detectors and smaller pixel sizes
on STIS allow a better estimate of the wavelength and spatial dependence of scattered light than the one-dimensional GHRS digicon science diodes. After the scattered light correction was applied to the STIS data, the residual flux in the core of the saturated H~{\sc i} profile is less than 1\% of the continuum flux at 1213 \AA.   The GHRS observations were done with 
the default STEP-PATT option, where the background is measured
with the science diodes for only 6\% of the total exposure time (Soderblom
1995). Due to the low S/N of this background measurement,  
only a low-order polynomial fit can be made to the background variations
(Cardelli et al. 1990). Use of the standard values of the four echelle
scatter correction coefficients recommended by Cardelli et al. (1993) yields a
significant ($\sim$ 8\% of the continuum flux at 1213 \AA) oversubtraction in the  core of the saturated H~{\sc i} profile (see Fig 1, VM98). We corrected for this by multiplying the background by 0.9 prior to subtraction ($\S$2), but if there are significant variations in the background with wavelength, this ad hoc correction is inadequate. Most previous D/H ratios derived with GHRS have used late-type stars as the background sources and these emission-line sources
should have a much less significant scattered light problem. Given the importance of the question of 
spatial variations of the D/H ratio in the LISM, future STIS observations of G191-B2B with a smaller aperture (to achieve higher wavelength resolution),
would be valuable. 
 \acknowledgements
This research was supported by a GTO grant to the STIS IDT. MAB acknowledges support from PPARC, UK. We are grateful to Jeff Linsky, the referee, for insightful and helpful comments.

\clearpage
\begin{deluxetable}{lcccccccr}
\small
\tablenum{1}
\tablecaption{RESULTS OF PROFILE FITTING \label{tbl-1}}
\tablehead {{Fit type} & \colhead{$\chi$$^2$/$\nu$\tablenotemark{}} 
& \colhead{$\nu$\tablenotemark{a}} & \colhead{} & \colhead{Comp 1} & \colhead{} 
& \colhead{} & \colhead{LIC Component} 
& \colhead{} \nl
\colhead{} & & & \colhead{N(H~\sc i)\tablenotemark{b}} 
& \colhead{$T$ \tablenotemark{c}}& \colhead{D/H\tablenotemark{d}}
& \colhead{N(H~\sc i)\tablenotemark{b}} 
& \colhead{$T$ \tablenotemark{c}} & \colhead{D/H\tablenotemark{d}}}
\startdata
STIS-FREE\tablenotemark{1} & 1.065 & 457 & 0.32$\pm$0.11 & 11110$\pm$820 & 2.24($>$ 1.26)\tablenotemark{\dagger} & 1.72$\pm$0.10 & 6130$\pm$830 & 1.60$\pm$$\stackrel{0.39}{_{0.27}}$\nl
\nl
STIS-FIXED\tablenotemark{2} & 1.065 & 458 & 0.41$\pm$0.04 & 10570$\pm$520 & 
1.71$\pm$$\stackrel{0.32}{_{0.24}}$ & 1.65$\pm$0.05 & 5740$\pm$800 & \nodata \nl
\nl
GHRS-FREE\tablenotemark{1} & 1.200 & 740 & 0.66$\pm$0.08 & 8590$\pm$630 & 0.66$\pm$$\stackrel{0.54}{_{0.05}}$ & 1.51$\pm$0.08 & 7090$\pm$780 &1.29$\pm$$\stackrel{0.21}{_{0.17}}$ \nl
\nl
GHRS-FIXED\tablenotemark{2} & 1.203 & 741 & 0.53$\pm$0.04 & 9040$\pm$510 & 1.17$\pm$$\stackrel{0.12}{_{0.11}}$ & 1.62$\pm$0.04 & 6900$\pm$770 & \nodata \nl
\nl
\enddata
\tablenotetext{a}{$\nu$ = number of degrees of freedom}
\tablenotetext{b}{in units of 10$^{18}$ cm$^{-2}$}
\tablenotetext{c}{Temperature in $K$ derived from H I thermal velocity dispersion}
\tablenotetext{d}{D/H in units of 10$^{-5}$; errors quoted denote 2$\sigma$ (95\%) confidence limits}
\tablenotetext{1} {Profile fitting keeping the value of D/H free in both velocity components}
\tablenotetext{2} {Profile fitting forcing the same value of D/H in both velocity components}
\tablenotetext{\dagger}{In the STIS-FREE profile fit, the region of the D~{\sc i} absorption
line that contributes most to the D~{\sc i} column density of comp 1 reaches close to zero flux (see Fig 2a), resulting in a D/H ratio that is not well constrained.}

\end{deluxetable}

%
%

\clearpage
%
%
\clearpage
\figcaption[fig1.ps]
{Normalized STIS echelle spectra of interstellar absorption lines in the line-of-sight to G191-B2B are plotted against heliocentric velocity for (a) N~{\sc i}
$\lambda$1200.7 (b) Si~{\sc ii} $\lambda$1304 and (c) Si~{\sc iii} $\lambda$1206.5. 
The data points are shown with $\pm$ 1$\sigma$ error bars. The best-fit models to the absorption line profiles are plotted as continuous lines. The dotted lines indicate the
normalized fluxes predicted by the WD NLTE stellar atmosphere models discussed in $\S$~3. The lower panel for each absorption line shows the residuals to the fits with the y-axis in each lower panel ranging from -0.1 to 0.1 in relative intensity units. Both N~{\sc i} and Si~{\sc ii} absorption lines are uncontaminated by WD photospheric lines and show two distinct velocity components (indicated by arrows), one at $v$$_{hel}$ $\sim$ 8.6 km s$^{-1}$ (comp 1) and another at $v$$_{hel}$ $\sim$ 19.3 km s$^{-1}$ (LIC component).
In (c), the Si~{\sc iii} GHRS profile is plotted as a histogram ($\pm$ 1$\sigma$ error bars plotted every $\sim$ 7 data points) over the STIS profile and shows the good agreement between the two datasets ($\S$~4).  
We are able to obtain excellent fits to both the STIS and GHRS profiles of Si~{\sc iii} 1206.5 \AA\ using only two velocity components. 
\label{fig1}}

\figcaption[fig2.ps]
{Profile fits keeping the value of
the D/H ratio free in both velocity components, obtained using
(a) STIS data and (b) GHRS data. The observed data are shown as histograms. $\pm$ 1$\sigma$ error bars are shown for every $\sim$ 25 data points. The best-fit models to the absorption line profiles are plotted as continuous thin lines and the lower panel for each absorption line shows the residuals to the fits. The individual component fit for the LIC absorption component is shown by dotted lines while comp 1 is shown by dashed lines
The dot-dashed line shows the intrinsic WD Lyman-$\alpha$ profile predicted by the line-blanketed stratified NLTE atmosphere models ($\S$3) which was used in the profile fitting.  In the case of the GHRS spectra (b), the region $\sim$ 1215.7 \AA\ which contains
the geocoronal feature, has been removed from the data before profile fitting. 
\label{fig2}}

\figcaption[fig3.ps]
{A comparison of the STIS (continuous histogram) and GHRS (dotted histogram)
data in the 1215 to 1216.5 \AA\ region. 
A 3-pixel smoothing has been applied to both data sets, and the zero levels and scale factors of both data sets have been adjusted to provide agreement with
theoretical models.  Note the D~{\sc i} line in the GHRS spectrum is shallower in comparison to the STIS spectrum. The D/H ratios for the two velocity components obtained from STIS data are consistent with (D/H)$_{LIC}$ = 1.5$\pm$0.1$\times$10$^{-5}$ obtained by Linsky (1998) and provide no evidence for a variation in the D/H ratio within $\sim$ 70~pc. \label{fig3}}

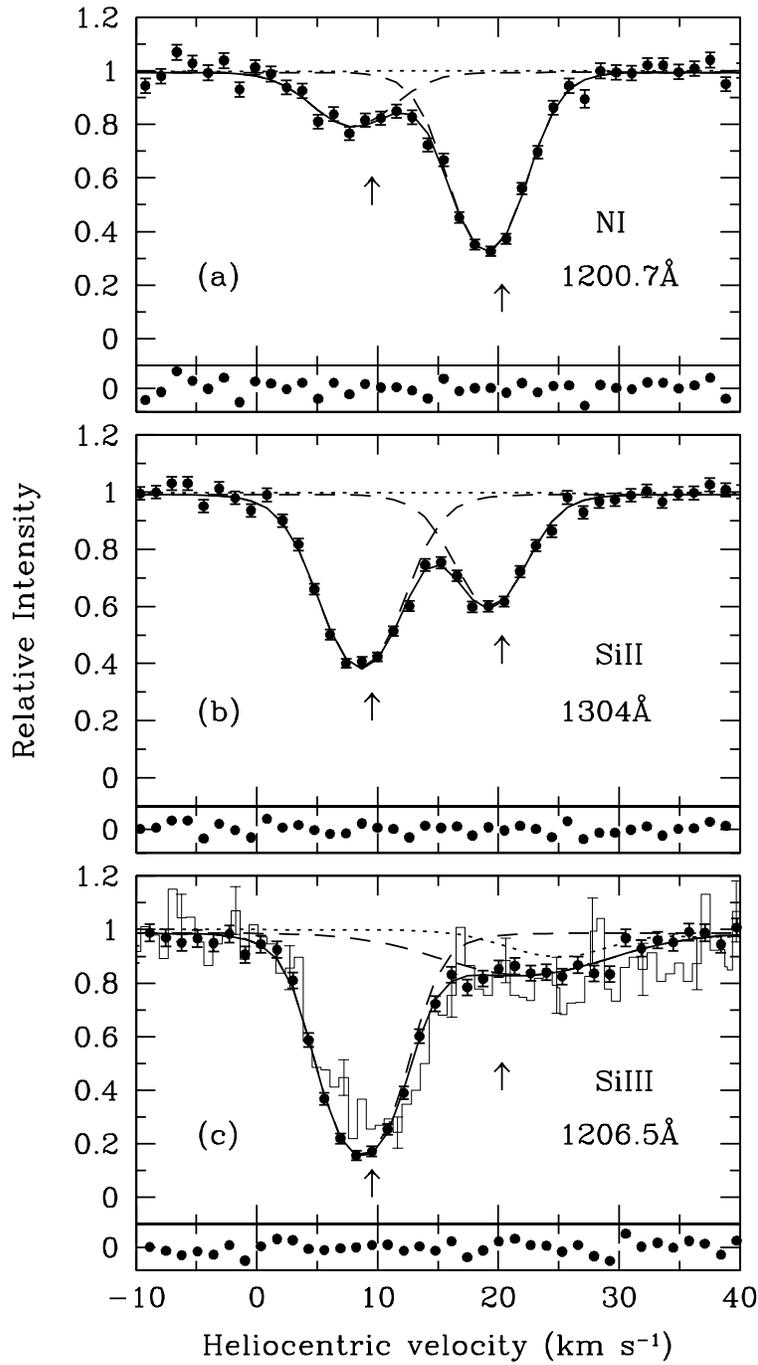
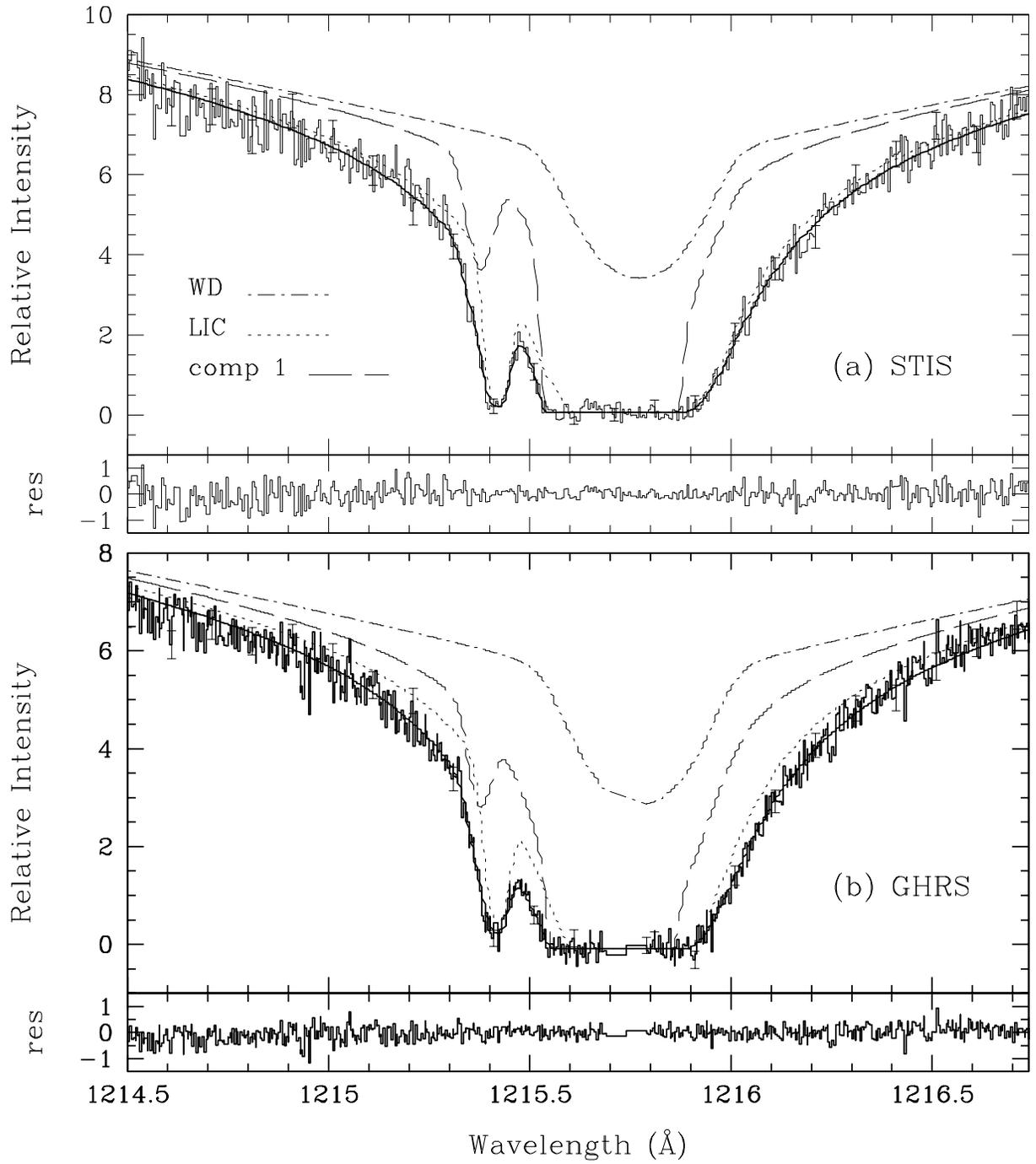
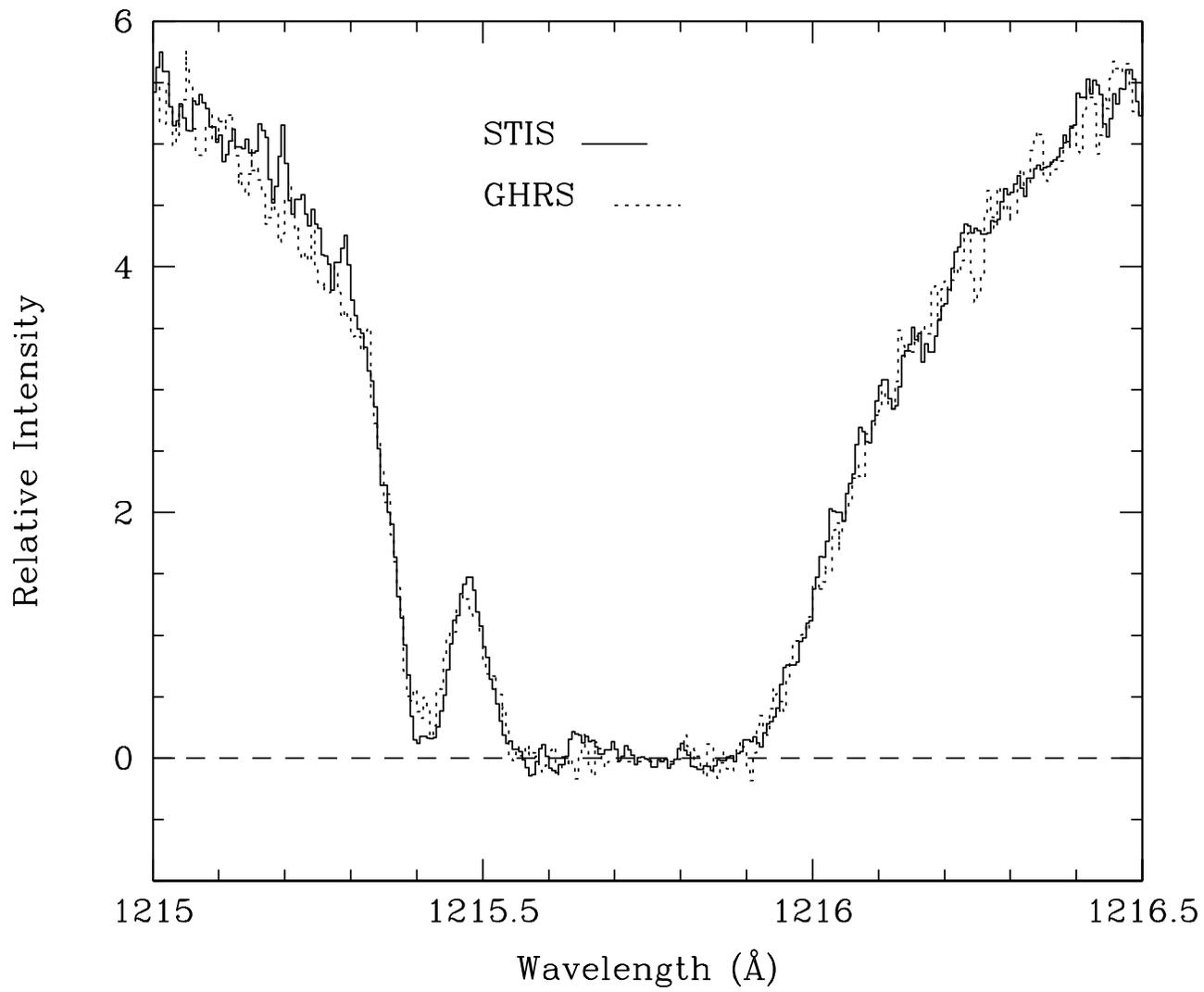
\end{document}